# Measuring the thermal conductivity and interfacial thermal resistance of suspended MoS$_2$ using electron beam self-heating technique

Adili Aiyiti[1,2,3*], Xue Bai[1,2,3,4*], Jing Wu[5], Xiangfan Xu[1,2,3], Baowen Li[6]

[1]Center for Phononics and Thermal Energy Science, School of Physics Science and Engineering, Tongji University, 200092 Shanghai, China

[2] China-EU Joint Center for Nanophononics, School of Physics Science and Engineering, Tongji University, 200092 Shanghai, China

[3]Shanghai Key Laboratory of Special Artificial Microstructure Materials and Technology, School of Physics Science and Engineering, Tongji University, 200092 Shanghai, China

[4]NUS Graduate School for Integrative Sciences and Engineering, National University of Singapore, Kent Ridge 119620, Republic of Singapore

[5]Institute of Materials research and Engineering (IMRE), 2 Fusionopolis Way, Innovis, #08-03, 138634, Republic of Singapore

[6]Department of Mechanical Engineering, University of Colorado, Boulder, CO 80309-0427, USA

* These authors contributed equally to this work.
Correspondence and requests for materials should be addressed to X.X. (email: xuxiangfan@tongji.edu.cn ) or to B.L. (email: Baowen.Li@colorado.edu ).

**Abstract**

Establishment of a new technique or extension of an existing technique for thermal and thermoelectric measurements to a more challenging system is an important task to explore the thermal and thermoelectric properties of various materials and systems. The bottleneck lies in the challenges in measuring the thermal contact resistance. In this work, we applied electron beam self-heating technique to derive the intrinsic thermal conductivity of suspended Molybdenum Disulfide (MoS$_2$) ribbons and the thermal contact resistance, with which the interfacial thermal resistance between few-layer MoS$_2$ and Pt electrodes was calculated. The measured room temperature thermal conductivity of MoS$_2$ is around ~ 30 W/mK, while the estimated interfacial thermal resistance is around ~ $2\times10^{-6}$ m$^2$K/W. Our experiments extend a useful branch in application of this technique for studying thermal properties of suspended layered ribbons and have potential application in investigating the interfacial thermal resistance of different 2D heterojunctions.

**1 Introduction**

Investigating the thermal and thermoelectric properties of a material is more intractable than the electrical properties, especially for the measurement of intrinsic thermal conductance or resistance of low dimensional systems. This may partially due to the special nature of phonons, which could leak to any contact linked with and can't be controlled as easily and precisely as electrons so far. Thus, when measuring the intrinsic thermal conductance or resistance of low dimensional systems, one has to suspend the samples, especially the 2D materials like graphene, to eliminate the influence of the supporting substrate [1, 2].

Nevertheless, quite some techniques have been established to investigate the thermal and thermoelectric properties of low dimensional systems. The utility $3\omega$ technique developed by Cahill et al. [3] to explore the thermal conductivity of thin film has been extended to study the thermal properties of other low dimensional systems, e.g. , thermal conductivity [4, 5], diffusivity, specific heat [4-6] of nanowires (NWs) [7-9] and carbon nanotubes (CNTs) [6, 10], interfacial thermal resistance [11], etc. In most cases, this technique was applied to investigate the through-plain thermal conductivity and interfacial thermal resistance of layered structures supported by substrates, which could confine the initial phonon modes in the samples. Recently, Raman thermometry-based techniques have been explored to probe thermal transport in CNTs [12], graphene [13], MoS$_2$ [14, 15], GaAs NWs [16] and porous Silicon [17], etc. Although it can be applied in both supported and suspended systems, the error caused by weak correlation between temperature and Raman shift in some systems could limit its applications. Besides, a non-contact pump-probe measurement method, the time domain thermoreflectance (TDTR) technique firstly demonstrated by Paddock and Eesley [18] has been extended to various novel and sophisticated measurements [19-26], including both in-plain and through-plain thermal conductivity of supported samples. Compared with the other techniques mentioned above, the TDTR technique requires more sophisticated and expensive setups as well as very careful operation to obtain data that need to be derived by extra modeling and analysis. Last but not least, to eliminate the confinement of the supporting substrates the suspended resistance thermometer microdevices fabricated by Shi and Kim et al. [27, 28] for thermal and thermoelectric measurements of NWs [29] have been extensively applied to measure the thermal properties of two dimensional (2D) materials including graphene [1], $MoS_2$ [30], boron nitride [31] and black phosphorous [32], etc. It is worth to noting that the first experimentally investigation of phononic devices including thermal diode [33] and thermal memory [34] was based on this technique. However, the inevitable thermal contact resistance that the samples encountered at the two ends is an intractable issue. Fortunately, Wang et al. [35] and Liu et al. [36] established a technique named focused electron beam self-heating technique based on suspended thermal bridge devices, in which the thermal contact resistance was elaborately eliminated, to derive the intrinsic thermal conductivity of supported graphene and suspended NWs. However, this technique facing challenges when applied to measure the suspended 2D Nano materials since it is very difficult to obtain clean and truly suspended polymer-free 2D samples and to squeeze reasonable signal and data from the limited energy absorption caused by limited interaction between incident electrons and sample atoms.

Here, we utilized the electron beam self-heating technique to directly derive the intrinsic thermal conductivity and thermal contact resistance of few-layer $MoS_2$ ribbons，which was prepared by dry transfer method. Both experimental and Monte Carlo simulation results suggests that a better signal could be achieved for a thicker layered ribbon that contains larger atoms by modifying the acceleration voltage and spot size of the incident electron beam of the scanning electron microscope (SEM).

## 2 Results and Discussion

Few-layer $MoS_2$ flakes, which were exfoliated from the bulk $MoS_2$ crystal by scotch-tape, were transferred onto the transparent solid-state Polydimethylsiloxane (PDMS) blocks. Some thinner and oblong flakes were chosen and carefully aligned to the prepatterned suspended thermal bridge devices with a homemade transferring platform under microscope. Sample preparation process is illustrated in Fig. 1(a-d). The prepatterned devices were cleaned by oxygen plasma and annealed at 225 °C in $H_2/Ar$ atmosphere for two hours to clean the possible residue on the top before the flakes are transferred. In contrast to the Polymethylmethacrylate (PMMA) mediated wet-transfer method, during the dry transfer method less polymer residues would contaminate the sample surfaces [31]. As prepared samples were annealed again in the same condition mentioned above to clean the possible residue on the samples and enhance the thermal contact between the sample and electrodes before any thermal measurement. The dimensionality including length, width as well as the surface morphology, edge profile of the samples were characterized by FEI SEM (Fig.1(e-g)), while the thickness of the samples was confirmed by Raman Spectroscopy [37] according to the peak positions and distances of $E^1_{2g}$ and $A^1_g$ (Fig.1(h)).

The samples were transferred into the SEM chamber for measurement of thermal resistance with the focused electron beam self-heating technique [36]. Fig. 2 demonstrates the schematic and measurement setups of this technique, where scanning electron beam is used as heating source while the two suspended membranes as temperature monitors. Slightly different from thermal bridge method, where one island acts as heater and the other acts as sensor, the identical designed two islands of the thermal bridge device are both regarded as sensors. The focused electron beam act as a noncontact localized heat source and is scanned along the length of the $MoS_2$ samples, during which part of the energy of the electron beam is absorbed by the local spot.

The energy loss of incident electrons is mainly from inelastic scattering arising from Coulomb interaction between the incident electron and the atomic electrons surrounding the nucleus. Base on Bethe's deduction of the rate of energy loss $dE$ with traveled distance $ds$ is [38]

$$\frac{dE}{ds}\left(\frac{\text{keV}}{\text{cm}}\right) = -2\pi e^4 N_0 \frac{Z\rho}{AE_i} \ln\left(\frac{1.166E_i}{J}\right) \tag{1}$$

$$J(\text{keV}) = (9.76Z + 58.5Z^{-0.19}) \times 10^{-3} \tag{2}$$

where $e$ is the electron charge, $N_0$ is Avogadro's number, $Z$ is the atomic number, $\rho$ is the density (g/cm$^3$), $E_i$ is the electron energy (keV) at any point in the specimen, and $J$ is the average loss in energy per event. Thus, $\frac{dE}{ds}$ in MoS$_2$ is dependent on the incident electron energy. Because the thickness of the suspended few-layer MoS$_2$ flakes is few nanometers, which is much smaller than the reported sample thickness by using the electron beam heating technique [35, 36], the absorbed energy will be small. The limited absorbed energy will lead a low temperature rise at two sensors and make it difficult to acquire reasonable signal. Therefore, to choose proper incident electron energy, a Monte Carlo simulation [39] has been conducted to illustrate the absorbed energy of different layers of MoS$_2$ samples under different incident electron beam energy. The schematic of the simulation model shown in Fig. 3a is drawn to visualize the simulation situation. Fig. 3b shows the simulation results. The energies absorbed by samples increases with the increasing thickness and decreases with the increasing incident energy. Thus, we have chosen a lower acceleration voltage (15 keV) to get a better signal.

The heat generated at the local spot flows apart towards the two sensor islands and rises the temperature of the sensors. The detailed derivation process of the relevant calculation equations was described in the previous works [35, 36, 40]. The cumulative thermal resistance ($R_i(x)$) from the left island to the heating spot is given by following equations:

$$R_i(x) = R_b \left[\frac{\alpha_0 - \alpha_i(x)}{1 + \alpha_i(x)}\right] \tag{3}$$

$$\alpha_0 = \frac{\Delta T_{L0}}{\Delta T_{R0}} \tag{4}$$

$$\alpha_i = \frac{\Delta T_L}{\Delta T_R} \tag{5}$$

where $R_b$ is the equivalent thermal resistance of the six suspension beams connecting the left (or the right) island to the environment and is derived from the thermal bridge method measurement. The temperature rises of the left and right island, $\Delta T_{L0}$ and $\Delta T_{R0}$, are also measured by thermal bridge method. $\Delta T_L$ and $\Delta T_R$ are corresponding temperature rise when the focused electron beam is scanned along the length of the samples. Thermal conductivity of the samples is calculated by following equation:

$$\kappa = \frac{1}{(dR_i/dx) \cdot A} \tag{6}$$

where $A$ is the cross-section area of the samples.

Fig. 4(a) shows the cumulative thermal resistance ($R_i$) as a function of the distance from the right sensor to the heating spot. The thicker (5$L$) sample (S1) shows relatively smooth $R_i$ curve (denotes better signals) compared to the thinner (4$L$) sample (S3). By contrast, the $R_i$ curve of narrower sample (S2) deviate from the others. Nevertheless, the slopes of the fitted lines to the three samples are almost the same. With the assumption that the thermal transport along the strip obeys Fourier's law [41-43], the thermal resistance of the entire strip is considered as that of the left and right segments in series, and is constant irrespective of the heat source position. Hence, the thermal conductivity of the sample is simply derived by substituting the slope of the linear fitting into the equation (6). As shown in Fig. 4(b), the thermal conductivity values derived from the focused electron beam self-heating technique agree with the ones that are measured with conventional thermal bridge method and Raman thermometry-based technique [15].

The derived room-temperature thermal conductivity of 4 layer $MoS_2$ (S2) is 34±6 W/mK which is comparable to the results in some previous publications [15, 44, 45]. However, our results are lower than that in some previous works, e.g. , (52 W/mK) [14] measured with Raman-based technique and (44-50 W/mK) [30] measured with suspended thermal bridge method. The lower thermal conductivity values in our experiment may originate from the different sample seeds, rough edges induced during transfer and from measurement method. Of course, partially due to the inevitable factors mentioned above, even the exact thermal conductivity value for other 2D materials including the most widely studied 2D material, graphene, has not been nailed down [2]. Phonons behave differently in nanostructures when samples are shrunk from bulk into 2D structures.

Of particular fundamental interest is the thermal conductivity dimensional crossover from 3D to 2D, as what has been observed in graphene. As the flexural phonon modes [46] in the graphene dominates the heat conduction, thermal conductivity increases with decreasing number of layers [47]. for the out-of-plane acoustic phonons encounters damping caused by inter-layer coupling. MD results reported by Ding et al. [48] show that the in-plane thermal conductivity of multilayer $MoS_2$ is insensitive to the number of layers, probably due to the fact that phonon-phonon scattering channel keeps unchanged with varying layer thickness and due to the strong inter-layer coupling in $MoS_2$ than that in graphene [42, 48]. This is in strong contrast to the in-plane thermal conductivity of graphene, in which the interlayer interaction strongly affects the in-plane thermal conductivity. However, first principle calculation results reported by Gu et al. [49] indicates that the basal-plane thermal conductivity of $MoS_2$ decreases with increasing number of layer. Interestingly, experimental evidence for the exact trend of layer dependency of thermal conductivity for $MoS_2$ is still under debate, e.g. Bae et al. [50] reported that thermal conductivity of $MoS_2$ increases with increasing layer number, while Yan et al. [15] reported 34±4 W/mK (1 layer); Zhang et al. [51] got 84±17 W/mK (1 layer) and 77±25 W/mK (2 layer); Jo et al. [30] measured 44-50 W/mK (4 layer) and 48-52 W/mK (7 layer); Sahoo et al. [14] obtained 52 W/mK (11 layer); Liu et al. [52] reported 85-112 W/mK (bulk). The existing experimental data combined together is still insufficient to nail down the exact trend of layer dependency of thermal conductivity for $MoS_2$ from the aspect of experiment. More systematic investigation on layer dependency of $MoS_2$ need to be done.

The total thermal resistance ($R$) can be measured utilizing thermal bridge method. As shown in Fig. 5, the measured thermal resistance is $3.5\times10^6$ K/W, $12.7\times10^6$ K/W and $13.4\times10^6$ K/W for S1, S2 and S3 respectively. $R\times W$ product is plotted as a function of $L/t$ in the inset for Fig. 4 and the beat linear fitting to the data is extended to the intercept value of the vertical axis where $L=0$ and $R\times W=R_c\times W$, by which the thermal contact resistance ($R_c$) can be calculated. The derived thermal contact resistance from the thermal bridge (TB) method is $1.7\times10^6$ K/W, $3.3\times10^6$ K/W and $2.65\times10^6$ K/W for the three samples respectively. As the total thermal resistance ($R$) measured with the thermal bridge method and intrinsic thermal conductivity derived from the focused electron beam (EB) self-heating technique are known, a set of thermal contact resistance can also be calculated, i.e. $1.8\times10^6$ K/W, $5.1\times10^6$ K/W and $3.6\times10^6$ K/W, which are larger than the ones derived from the thermal bridge method. Furthermore, the interfacial thermal resistance ($R_{\text{interface}}$) (ITR) can be estimated from the equation [53] below:

$$\frac{R_c}{2} = \left[\sqrt{\frac{\kappa A W}{R_{\text{interface}}}} \tanh\left(\sqrt{\frac{W}{\kappa A R_{\text{interface}}}}\, l_c\right)\right]^{-1} \qquad (5)$$

, where $\kappa$ is thermal conductivity, $R_c$ is thermal contact resistance, $A$ is contact area connecting the sample and electrode and $W$ is the width of the sample. The estimated interfacial thermal resistance between few-layer $MoS_2$ and Pt electrode is ~ $2.4\times10^{-6}$ m$^2$K/W, ~ $1.8\times10^{-6}$ m$^2$K/W and ~ $1.9\times10^{-6}$ m$^2$K/W respectively, which are slightly larger with the ITR of 7 layer $MoS_2$ and crystalline silicon ( ~ $1\times10^{-6}$ m$^2$K/W); [54] single layer $MoS_2$ and $SiO_2$/Si substrate ( ~ $0.5\times10^{-6}$ m$^2$K/W) [55] measured with Raman spectroscopy; multilayer graphene and $SiO_2$/Si substrate ( ~ $0.6\times10^{-6}$ m$^2$K/W) [56] monitored with enhanced opto-thermal method. The lager ITR in our measurements is probably related to the interfacial contact condition between $MoS_2$ and Pt electrodes, whose interfacial roughness is larger.

Distinguished from the smooth $R_i$ curves in the previous nanowire [36] and SiN$x$ beam supported graphene [35], our data are not smooth. The nanowires are robust and pretty thick, while the graphene is supported by thick SiN$x$. Thus, there are relatively long paths for the high energy focused electron beam to go through and more atoms and electrons to interact with, which introduces more heat in the samples and increase the SNR (Signal to Noise Ratio). By contrast, our samples are more like suspended nets and most of the incident electrons pass through them without any interaction. The absorbed energies increased with increasing of the thickness and electron beam spot size, but decreasing of working voltage. Anyway, little amount of energy is absorbed by the samples. This is because only very limited amount of incident electrons interacts with the atoms and electrons of the suspended $MoS_2$ and heat up the local spot. One has to choose a lower acceleration voltage and a larger electron beam spot size to achieve a better signal, which sacrifices the special resolution of the measurement. And it requires high sample quality, especially the surface condition of 2D ribbons. Hence, it is very difficult to obtain reasonable signal for very thin samples due to the limited energy absorption from the incident electron beam. Besides, it requests more sensitive setups and is rather hard to obtain reasonable signal and data in the focused electron beam self-heating technique. Nevertheless, Intrinsic thermal conductivity of truly suspended 2D ribbons could be measured with the contact ignored. Thermal resistance, thermal contact resistance and interfacial thermal resistance could be derived, while the conventional thermal bridge method[57] have difficulty in that and the ITR data of $BN/SiO_2$ and $Graphene/SiO_2$ was used to estimation.

The fluctuant $R_i$ curves in the electron-beam self-heating technique could reflect the sample quality, including surface conditions and edge roughness. For instance, a lager $R_i$ values and obvious fluctuation with lager error bars are observed in the S2 compared with the others. A few possible sources that may cause the larger error bars should be taken into consideration. As we mentioned in the manuscript, it is different from the nanowire [36] and $SiN_x$ beam supported graphene [35] cases. The suspended few layer $MoS_2$ ribbon is more like a net that is bridging the microdevices, large amount of the incident electrons easily path through the sample without sufficient interaction with the sample atoms. The limited amount of energy absorption results from the insufficient interaction, which further results in weak signal in the signal monitoring and acquisition system. Besides, the focused electron beam may be out of focus somehow when scanning on the rippled surface of the sample, which may also disturb the low signal on some level. And the disturbance caused by the rough edges and amorphous carbon may also affect the measurements. Given this, there are some possible ways to further optimize the signal and reduce the error bar, i.e. prepare very high-quality samples with flat and clean enough surface and smooth edge; optimize the parameters and status of the focused electron beam; update the signal and data acquisition system with a more sensitive, stable and high-precision signal processor and amplifier.

3 Conclusion

In summary, the focused electron beam self-heating technique could be applied for the suspended layered ribbons, which are thick enough to provide adequate atoms and path for the incident high-energy electrons to interact with. The amount and efficiency of energy absorption by the lattice from the focused electron beam is an important factor to determine the availability of this technique for layered thin film ribbons. To get a more reasonable result, one should choose an optimized acceleration voltage and spot size for the focused electron beam with the compromise of spacial resolution of the thermal measurement. With sensitivity and accuracy to be further improved, this technique would find application in investigating the ITR of various 2D heterojunctions in the future.

**Acknowledgements**

This work is supported by the National Natural Science Foundation of China (No. 11674245 & No. 11334007), by Shanghai Committee of Science and Technology in China (No. 17142202100 & No. 17ZR1447900). J.W. is supported by A*STAR Pharos Funding from the Science and Engineering Research Council of Singapore (Grant No. 152 72 00015).

**Conflict of Interest**

The authors declare no conflict of interest.

**Figures and captions**

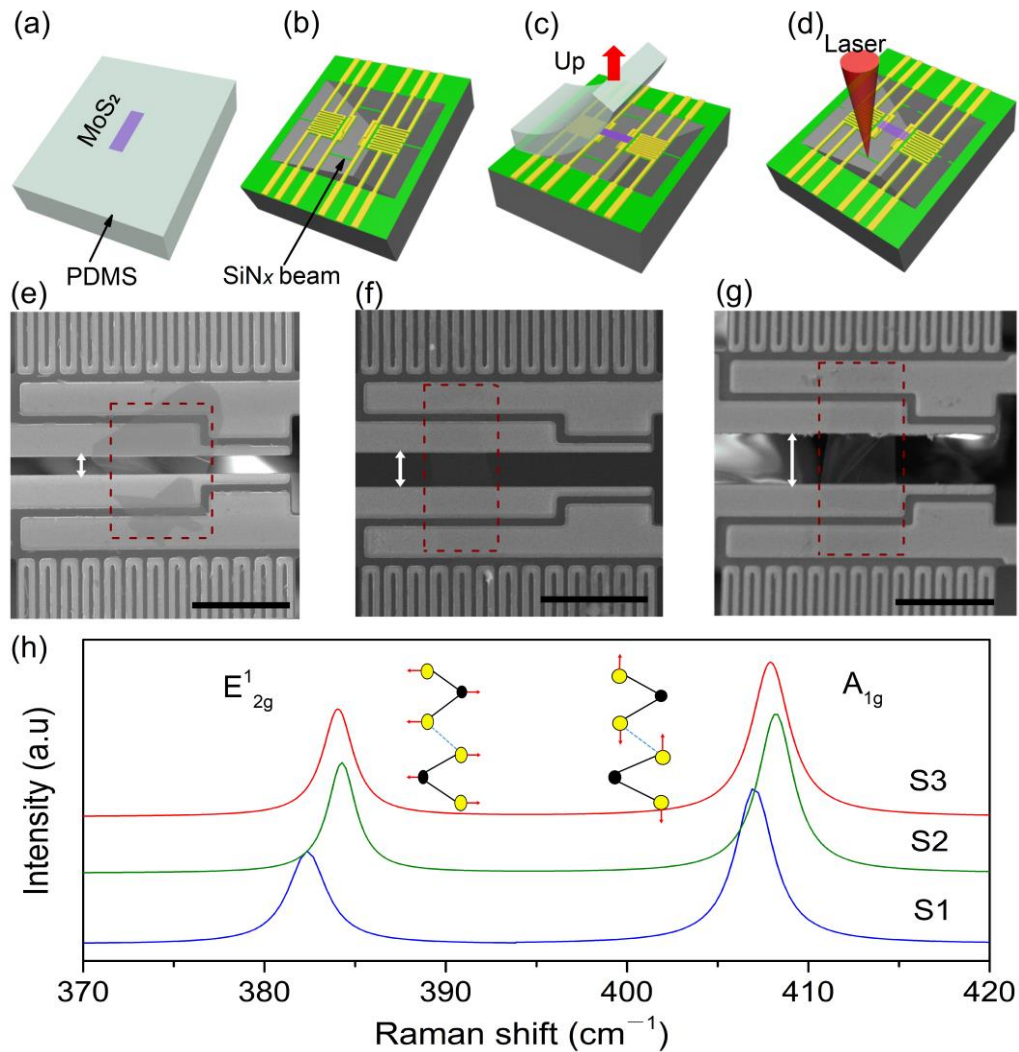

**Figure 1. Sample preparation and characterization.** (a) A long strip of $MoS_2$ is exfoliated on the transparent PDMS block. (b) Schematic of a prefabricated suspended thermal bridge device. SiN$x$ beams are designed to protect the device from damage during the wet etching process and dry transfer process. (c) The sample is aligned and transferred to the suspended device under the optical microscope before the PDMS block is stripped away. (d) The SiN$x$ beams are cut by laser beam. (e-g) SEM images of the prepared samples. The scale bar is 5μm. (h) Raman spectrum of the samples. The frequency difference between two Raman bands of $MoS_2$ ($E^1_{2g}$ and $A^1_g$) is 24.6 cm$^{-1}$ and 23.8 cm$^{-1}$ for 5$L$ (S1) and 4$L$ (S2 and S3) $MoS_2$, respectively.

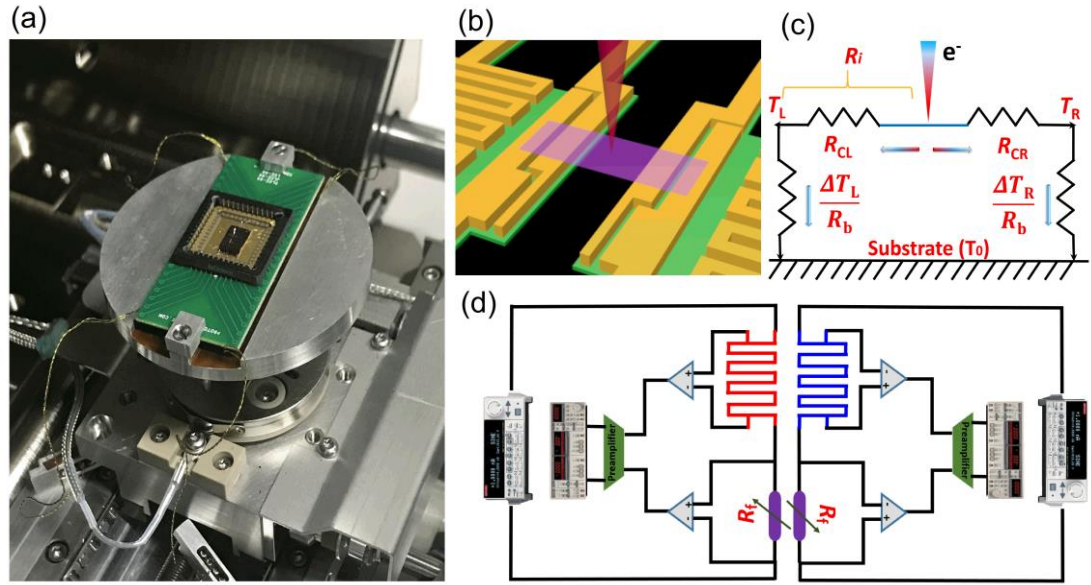

**Figure 2. The experimental setups and method.** (a) Picture of the measurement stage and chip carrier adapted on the SEM sample holder. (b) Schematic diagram of the focused electron beam heating technique. (c) The equivalent thermal resistance circuit for the focused electron-beam heating method, showing the $R_i$ from the left sensor to the local heat spot, the thermal contact resistance ($R_{CL}$ and $R_{CR}$), temperature rise of the left ($\Delta T_L$) and right ($\Delta T_R$) sensor. (d) Schematic of the measurement setups. The focused electron beam is scanned along the sample and the temperature rise of the both sensors are monitored. A differential offset compensation electrical circuit is applied to improve the sensitivity of the sensors.

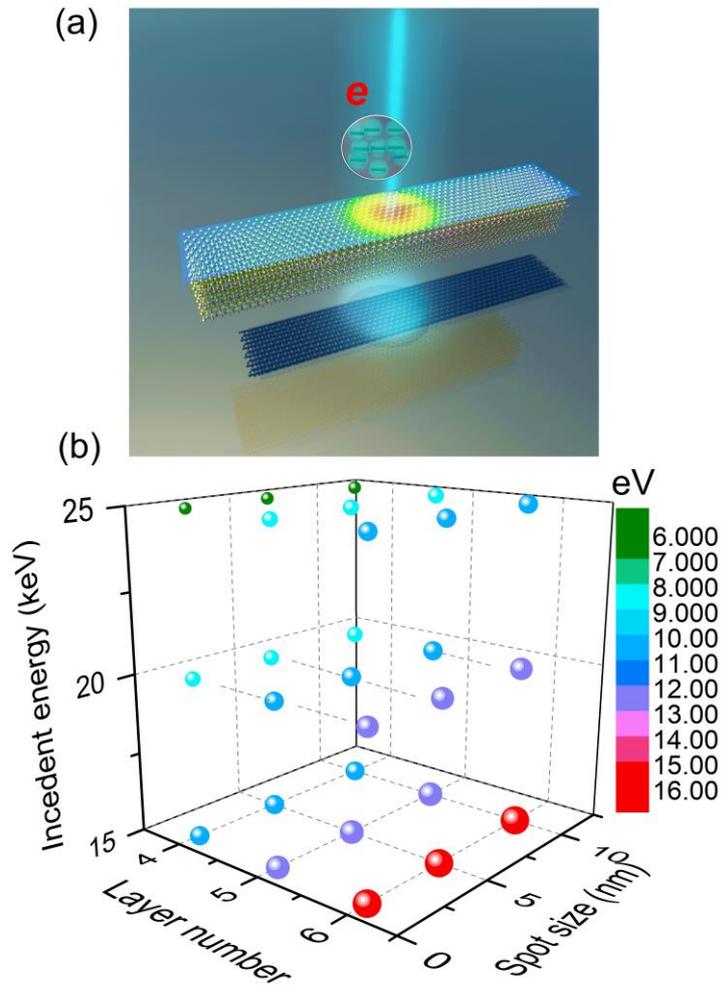

**Figure 3. Monte Carlo simulation.** (a) Schematic of simulation model. (b) Monte Carlo simulation of energy absorption for 4-6 layers $MoS_2$ samples with 1, 5, 10 nm electron beam diameters under 15, 20, 25 keV electron beam energy. The bigger symbol filled with deeper color denotes a larger energy absorption by the local spot where the focused electron beam shines on. In the experiment, 15 keV and 10 nm of electron beam is applied for the S2 and S3, while 15 keV and 5 nm of electron beam is applied for the S1.

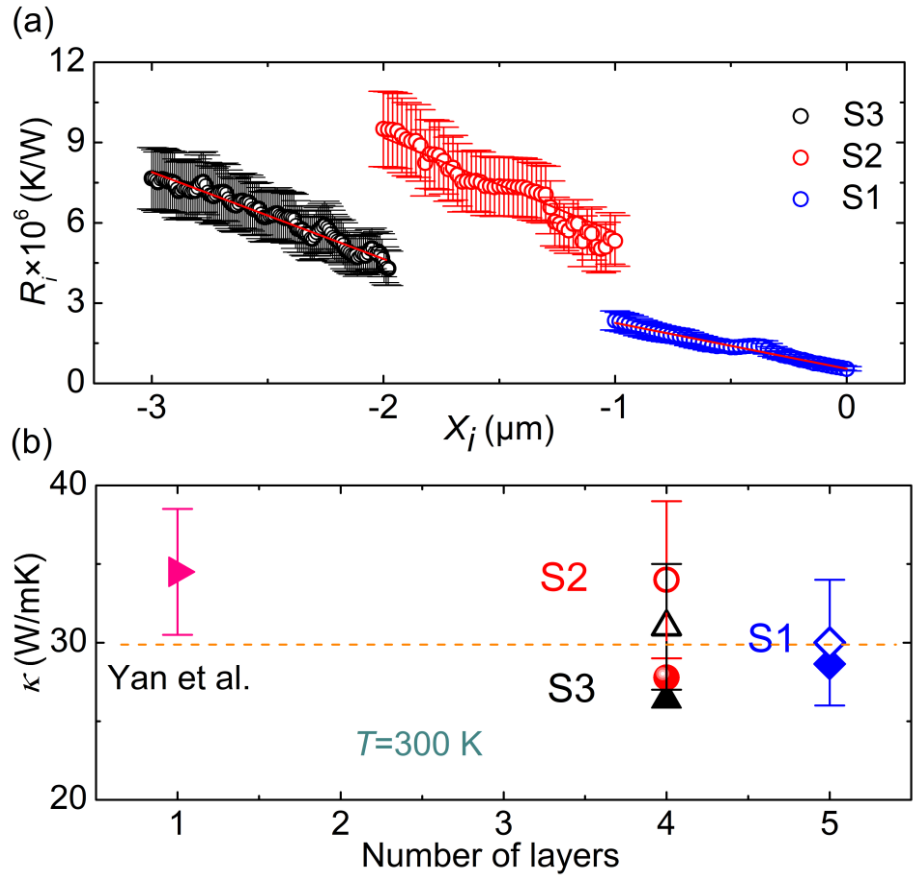

**Figure 4. Derivation of the intrinsic thermal conductivity.** (a) Cumulative thermal resistance ($R_i$) as function of a selected distance from the right sensor. Three different unfilled symbols denote three different samples. (The blue, red and black circles denote S1, S2 and S3 respectively.) For the sake of uniformity, the scanning window of SEM is fixed at ~ 3 μm for the samples. The solid lines are the best linear fitting to the data. (b) $\kappa$ vs. Number of layers. Filled symbols (blue, red and black) denote the thermal conductivity measured by thermal bridge method, while unfilled symbols denote the thermal conductivity derived from the focused electron beam self-heating technique. The experimental result (pink triangle) reported by Yan et al. [15] is given to compare with our results.

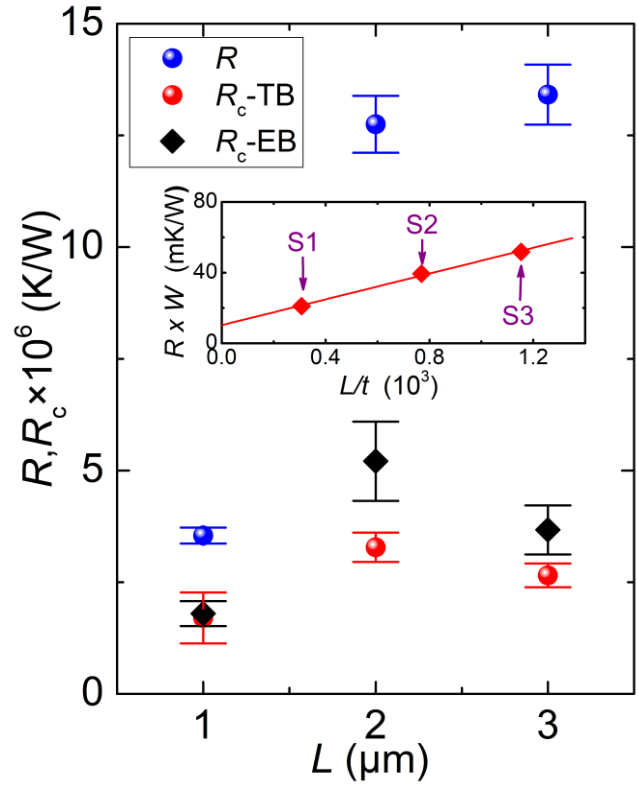

**Figure 5. Thermal resistance ($R$) and thermal contact resistance ($R_c$) vs. sample length ($L$).** Blue balls denote the thermal resistance of the samples; Red balls denote the thermal contact resistance of the samples derived from the data (Inset) obtained from the thermal bridge (TB) method, while the black diamonds denote the ones obtained from the electron beam (EB) self-heating technique. Inset: the $R \times W$ product vs. the $L/t$ ratio at $T=300$ K for the three samples. The red line is the best linear fitting to the data.